\documentclass[nofootinbib,a4paper,aps,pra,showpacs,preprintnumbers]{revtex4-2}
\usepackage[T2A]{fontenc}
\usepackage{amssymb}
\usepackage{amsmath}
\usepackage{mathtools}
\usepackage{braket}
\usepackage{tikz}
\usepackage{graphicx}
\usepackage{epstopdf}
\usepackage[utf8]{inputenc}
\usepackage[russian, english]{babel}
\usepackage{url}
\usepackage{listings}
\usepackage{gensymb}
\usepackage{xcolor}
\usepackage{ulem}
\usepackage{enumitem}
\usepackage{comment}

\newlist{steps}{enumerate}{1}
\setlist[steps, 1]{label = Step \arabic*:}

\begin{document}

\title{Quantum Algorithm for Researching the Nearest (QARN)}%

\author{Karina Zakharova\\
          reshetova.carina@yandex.ru}

\begin{abstract}
The search task is one of the most difficult when it comes to execution speed, and reducing the latter is important both when working with large data and with small samples, if they need to be processed frequently and in a limited time. Grover's algorithm gave hope to quantum computing and served as an excellent base for all possible implementations and modifications. In this paper, we propose a slightly different algorithm that increases the probability of finding the nearest value by reducing the probability of undesirable values in a controlled manner (in proportion to their difference from the desired value), as well as implementing an oracle that requires a single call without an additional ancilla qubit to redistribute the amplitudes.
\end{abstract}
\pacs{}

\maketitle

\hyphenation{}


\section{Introduction}
Quantum computing has been fascinated by the prospect of using quantum effects for decades \cite{1YO}. In particular, Shor's algorithm, based on the quantum Fourier transform, demonstrates the use of a qubit phase as an additional source for storing information \cite{2Shor}, \cite{3Copp}. The same technique is implemented in the phase estimation algorithm \cite{4Kit} and the algorithm for solving systems of linear equations, known as HHL \cite{5HHL}. These and other quantum algorithms have their own field of application and related limitations, but the tools implemented in their schemes can be combined and used autonomously to solve different problems: the use of phase, superposition, qubit entanglement, transition to multilevel qudits \cite{1YO}, \cite{6Amit}, \cite{7Wang}, \cite{8Mogos}. All these are powerful tools of quantum technologies, allowing one to obtain advantages in comparison with classical computations.
As for the data search problem, the most well-known approach to its optimization contains Grover's algorithm \cite{9Grov}. However, even here we have to face a number of difficulties in implementation, such as determination of the number of calls to this algorithm, which affects its accuracy; the number of possible solutions, which must also be taken into account, and also the presence of a hypothetical oracle function, or black box (also used in the Deutsch-Jozsa algorithm and the so-called Simon problem \cite{10Jozsa}, \cite{11Simon}), whose action is described only in the abstract. This scheme is an excellent illustrative example of quantum superiority and is often used as a demonstration and training material.
As mentioned above, Grover's algorithm searches for the number of the desired element in a data array and works with the counter $K$, which has a range of values from $k_0 = 0$ to $k_{m-1} = 2^{m}-1$, where m is the number of bits of the element in the array. This method implies that at least one element that exactly matches the sought one must exist, and the number of oracle queries depends on the number of required elements in the array. In practice, if the very existence of the value exactly coinciding with the desired solution is not guaranteed, it is necessary to find the nearest and most suitable solution in the database with a minimum error. In addition, a calibrable constraint provided by the algorithm is necessary to avoid unnecessary iterations that can worsen the result. For this purpose, QARN is proposed and contains the following actions with a description of the oracle implementation:
\begin{enumerate}
\item
Obtaining as input an array $A$ of m elements of size $n$ and a reference value $B$ of size $n$.
\item
Creating, with the help of quantum logic gates, a superposition of all $m$ elements of the array $A$ in a special quantum register of $n$ qubits and counter of $l$ $d$-level qudits. The qubits serve as a buffer for copying and storing the array $A$ and occupy a memory size equal to one element of the array being copied. The qudits act as an auxiliary tool to create a superposition: each state of the qudits is entangled with a single copy element of the array $A$.
\item
The bitwise implementation of finding the element nearest to $B$ occurs simultaneously over all $m$ elements, and also allows us to ignore the influence of the sign of the calculated difference. The result of comparing each element is recorded as a change in the probability of getting a qudit state entangled with the given element: the smaller the element matches the search conditions, the smaller the probability of getting a qudit state entangled with this element when measured.
\item
The measurement of the counter qudits. The resulting state will indicate the number of the elements that are searched for. 
\end{enumerate}

\section{The essence of the problem}
Let there exist an array $A = [A_0, A_1, \dots, A_{m-1}]$ and some value $B$, for which it is required to find in the array $A$ either an exact match or the nearest. Obviously, you should subtract $B$ from each element in turn, fix the value of difference at each step, at the end choose the smallest difference, and by its index refer to the element of the array $A$, taken by this way the nearest to the reference value $B$. In the classical version of the calculations, it is necessary to create a local copy $A’$ for array $A$ of identical size. When searching for the smallest difference, it would be necessary to perform successive calculations on each element of the copy and, in addition to that, to take into account the sign of the difference, which requires additional resources in the form of time.

\section{Execution of QARN}
\begin{figure}
\begin{center}
\includegraphics[width=1\linewidth]{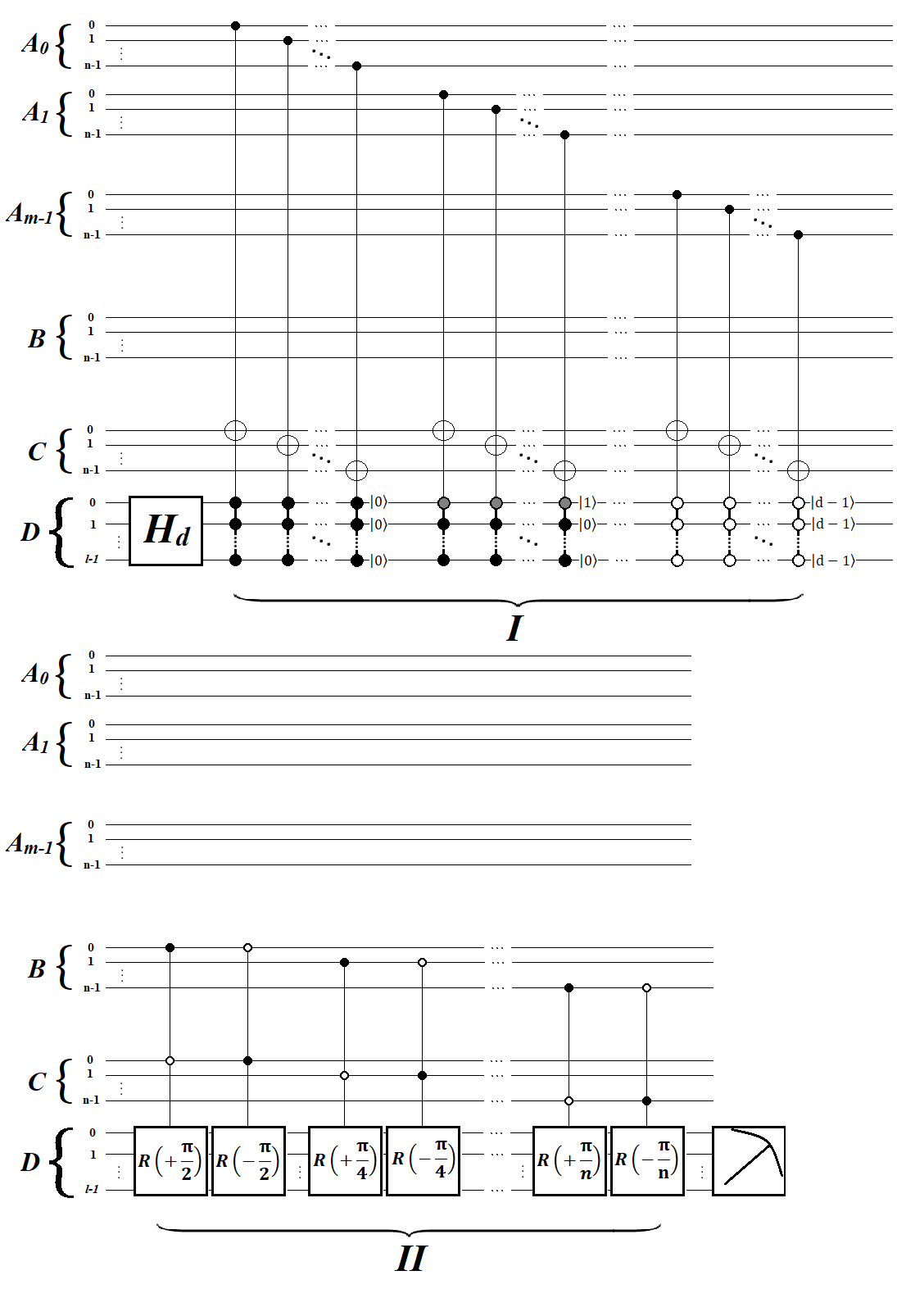}
\caption{\label{fig:1GenSch}General Scheme of QARN.}
\end{center}
\end{figure}
Figure~\ref{fig:1GenSch} shows a schematic diagram, implemented on a quantum computer, for finding the nearest value (to a given $B$) in an array $A$, where each of $m$ elements consists of $n$ qubits. The reference value $B$ also consists of $n$ qubits.
Let us assume that we need to copy data into the register of a quantum processor and do this explicitly while simultaneously confusing it with the counter. We will also assume that, as in the classical case, we do not have to work with the source information but only with its copy, so we recognize copying each element as the necessary classical part, after which we proceed to the quantum part, where, unlike the classical one, the entire array will be represented in a superposition in a single register of the quantum processor.
To create a local copy of the array $A = \sum_{i=0}^{m-1}A_i $, we will use the quantum register $C$ and counter $D$, where the register $C$ occupies $n$ qubits and counter $D$ contains $l$ $d$-level qudits ($D_0, D_1, \dots, D_{l-1}$). The number of possible states of the counter $D$ is proportional to the array $A$, i.e., $d^l = m$.
All qubits and qudits of the register and counter are initiated by zeros, the memory blocks $A_0\dots A_m$ and $B$ receive at input the values which are required to be processed by the task condition.
The first step creates a superposition of all possible states in counter $D$ by means of Hadamard gates, each of which is described by the formula \cite{12Carretta}:
\begin{equation}             \label{eq1}
 \begin{aligned}
 H_d = \frac{1}{\sqrt{d}} \sum_{\mu,\nu=1}^{d} e^{i 2\pi/d(\mu-1)(\nu-1)} |\mu\rangle\langle\nu|
  \end{aligned}
\end{equation}
Thus a superposition of all states is formed in the counter $D$:
\begin{equation}             \label{eq2}
 \begin{gathered}
D = \frac{1}{d^{\frac{1}{2}}}(|0_{l-1}\dots 0_1 0_0\rangle+ |0_{l-1}\dots0_1 1_0\rangle+\dots
\\
\dots+|(d-1)_{l-1}\dots(d-1)_1 (d-1)_0\rangle)
  \end{gathered}
\end{equation}
For convenience, each of the states $D$ will be denoted by $d’_j$, $j\in\{0\dots d_{l-1}\}$.
The first step is necessary to entangle one element of a copy of the array $A$ (placed in the register $C$) with each of the states obtained from the counter $D$. Which is why $d^{l}$ must be equal to $m$.
The entanglement of the qubits is carried out in step $I$ (the upper half of the Figure~\ref{fig:1GenSch}) by means of controlling gates, where the controlling states are the values $|1\rangle$ of the qubits of element $A_k$ and the states $d’_k$ of the counter $D$ such that $C_k$ (the $k^{th}$ superposition state of register $C$) becomes equal to $A_k$:
\begin{equation}             \label{eq3}
 \begin{gathered}
|B\rangle|A_0\rangle|A_1\rangle\dots|A_{m-1}\rangle|0_0 0_1\dots0_{n-1}\rangle|0_0 0_1\dots0_{l-1}\rangle
\\
\overset{H_d^{\otimes d}}{\rightarrow}
\\
\frac{1}{2^\frac{n}{2}}|B\rangle|A_0\rangle|A_1\rangle\dots|A_{m-1}\rangle|0_0 0_1\dots0_{n-1}\rangle
\\
(|0_0 0_1\dots0_{l-1}\rangle+|0_0 0_1\dots1_{l-1}\rangle+|d_0 d_1\dots d_{l-1}\rangle)
\\
\overset{I}{\rightarrow}
\\
\frac{1}{2^\frac{n}{2}}|B\rangle|A_0\rangle|A_1\rangle\dots|A_{m-1}\rangle(|A_0\rangle|0_0 0_1\dots0_{l-1}\rangle+
\\
+|A_1\rangle|0_0 0_1\dots1_{l-1}\rangle+\dots+|A_{m-1}\rangle|d_0 d_1\dots d_{l-1}\rangle)
  \end{gathered}
\end{equation}

\begin{figure}[h]
\begin{center}
\includegraphics[width=0.5\linewidth]{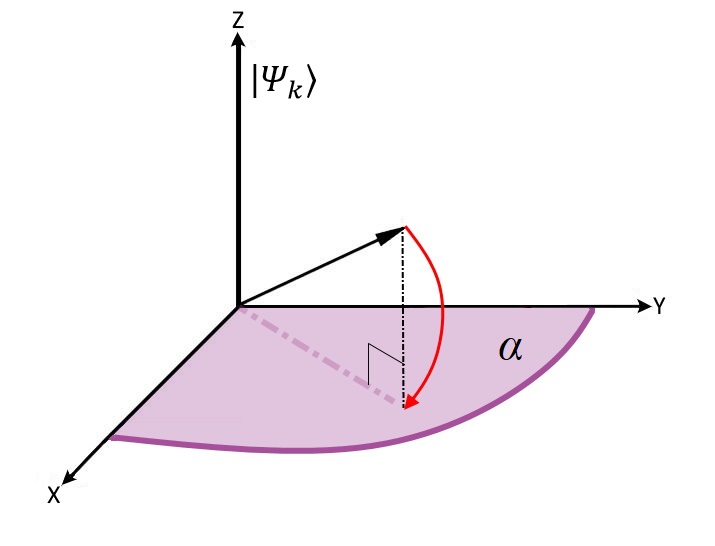}
\caption{\label{fig:2RotAbst}Rotation of the desired vector relative to the orthogonal hyperplane (the sum of all other vectors of the system states).}
\end{center}
\end{figure}

In step $II$ (the lower half of Figure~\ref{fig:1GenSch}), a sequence of control gates making turns in the counter $D$ is invoked. This part, in the "Grover" sense, is an oracle that will need a single call, as will be seen later. Here, the main task is to find the closest value by calculating the difference between $B$ and each element of the array $A$ stored in the register $C$. In doing so, the probability amplitude for each state $d’_j$ entangled with $C_j$ decreases proportionally to the difference $B – C_j$ obtained.
An abstraction of the action described is shown in Figure~\ref{fig:2RotAbst}.
If we represent all $N = 2^n$ possible states of some system as basic orthogonal vectors of the hyperspace of states, the superposition vector will be the result vector equidistant from the basic ones, respectively, when measuring this system the probability of getting each of its possible states is the same and equal to $1\frac{2}{\frac{n}{2}}$:
\begin{equation}             \label{eq4}
 \begin{gathered}
|\Psi\rangle=\frac{2}{\frac{n}{2}}(|0_0 0_1\dots0_{n-1}\rangle+|0_0 0_1\dots1_{n-1}\rangle+\dots\\
\dots+|1_0 1_1\dots1_{n-1}\rangle)=\sum_{i=0}^{N-1}|\Psi_i\rangle
  \end{gathered}
\end{equation}
The weakening of the probability of the $k^{th}$ state $|\Psi_k\rangle$ means the shift of the result vector $|\Psi\rangle$ towards the hyperplane $\alpha$ of all other basis vectors $(\sum_{i=0}^{N-1}|\Psi_i\rangle -|\Psi_k\rangle)$ and away from the vector $|\Psi_k\rangle$.
The difference $B – A_k = B – C_k$  is written as a rotation in counter $D$ using rotation gates such that the probability of a state $|D_k\rangle$ entangled with $|A_k\rangle$ decreases in proportion to the value of $B – A_k$. The comparison is bitwise, so the $i^{th}$ bit of the $n$-bit value of $B$ is subtracted from the $i^{th}$ bit of the $n$-bit register $C$. The value of the rotation is discrete, depends on the significance of the bits being compared and is equal to $\frac{\pi}{2^{i+1}}$: the difference of the higher bits is written as a rotation by $\frac{\pi}{2}$, the next bits by $\frac{\pi}{4}$, etc. The difference of the low bits will rotate the state of the counter $D$ by $\frac{\pi}{2^n}$.
As mentioned above, due to the superposition of states $\sum_{j=0}^{m-1}|A_j\rangle)$ concentrated in register $C$, the difference $B – A_j$ is counted simultaneously over all elements. At the end, counter $D$ is measured, and the resulting state entangled with a particular element of the array will point to that sought element. A general view of the space rotation matrix for the attenuated probability amplitude $C_j$ is shown below:
\begin{equation}             \label{eq5}
\left(\begin{array}{cccc}
\cos \frac{\varphi}{2} & \pm \frac{\sin \frac{\varphi}{2}}{\sqrt{2^n-1}} & \cdots & \pm \frac{\sin \frac{\varphi}{2}}{\sqrt{2^n-1}} \\
\pm \frac{\sin \frac{\varphi}{2}}{\sqrt{2^n-1}} & \cos \frac{\varphi}{2} & \cdots & \pm \frac{\sin \frac{\varphi}{2}}{\sqrt{2^n-1}} \\
\vdots & \vdots & \ddots & \vdots \\
\pm \frac{\sin \frac{\varphi}{2}}{\sqrt{2^n-1}} & \pm \frac{\sin \frac{\varphi}{2}}{\sqrt{2^n-1}} & \cdots & \cos \frac{\varphi}{2}
\end{array}\right)
\end{equation}

\section{Example for 2-level qudit (i.e. qubit)}
\begin{figure}
\begin{center}
\includegraphics[width=1\linewidth]{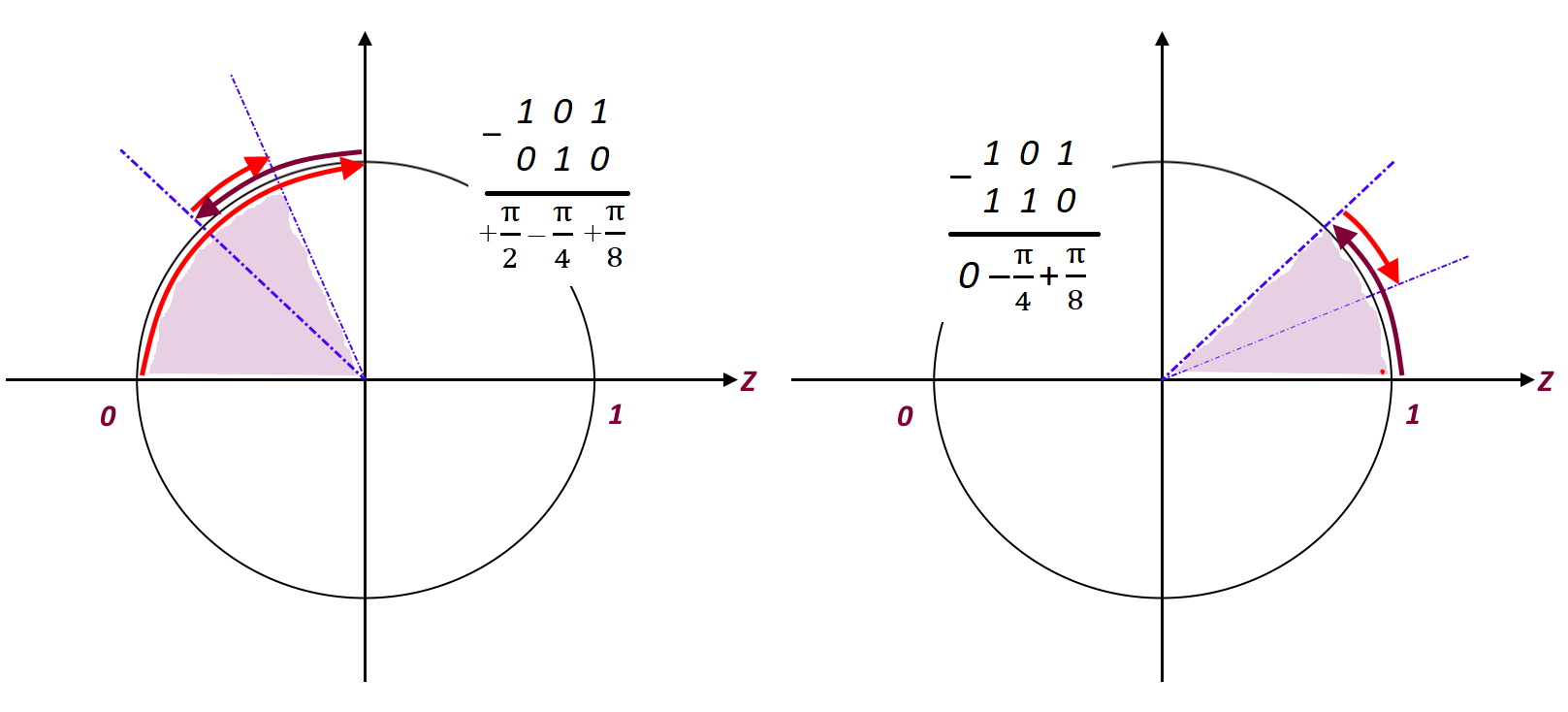}
\caption{\label{fig:4RotExmp}Abstraction of projection of subtraction into rotations.}
\end{center}
\end{figure}

Figure~\ref{fig:3ExmpSch} details a scheme for the case of $n=3$, $m=2$ and a two-level qudite $D (d = 2)$ corresponding to a classical qubit.
Displayed equations should be centered like in the example given below:
\begin{figure}
\begin{center}
\includegraphics[width=1\linewidth]{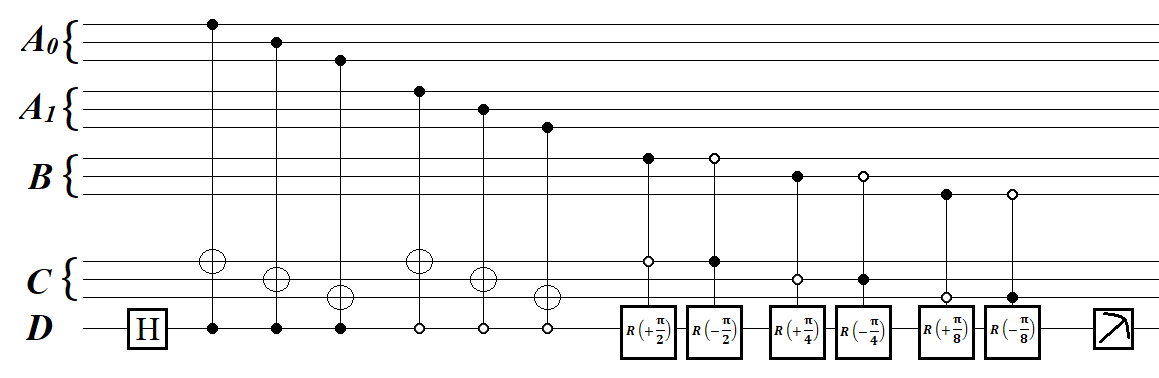}
\caption{\label{fig:3ExmpSch}Example of a circuit for an array of two 3-qubit elements.}
\end{center}
\end{figure}
The superposition creation function is a sequence of Toffoli and Pauli $X$ gates that rotate the state of the qubit around the $X$ axis by $\pi$:
\begin{equation}             \label{eq6}
 \begin{gathered}
R_X(\Theta)=\left(\begin{array}{cccc}
\cos\frac{\theta}{2}&-i\sin\frac{\theta}{2}\\
-i\sin\frac{\theta}{2}&\cos\frac{\theta}{2}
\end{array}\right)
  \end{gathered}
\end{equation}
The transformations after step $I$ are as follows:
\begin{equation}             \label{eq7}
 \begin{gathered}
|B\rangle|A_0\rangle|A_1\rangle|000\rangle|0\rangle
\\\overset{H}{\rightarrow}\\
\frac{1}{\sqrt{2}}|B\rangle|A_0\rangle|A_1\rangle|000\rangle(|0\rangle+|1\rangle)
\\\overset{I}{\rightarrow}\\
\frac{1}{\sqrt{2}}|B\rangle|A_0\rangle|A_1\rangle(|A_0 \rangle|0\rangle+|A_1\rangle|1\rangle)
  \end{gathered}
\end{equation}
where the coupling $CD$ will have state $\frac{1}{\sqrt{2}}(|A_0\rangle|0\rangle+|A_1\rangle|1\rangle)$. Then it is necessary to determine $|A_j\rangle$ nearest to $|B\rangle$ and strengthen the probability to get the state $A_j\rangle|d_j\rangle$ when measuring counter $D$ by writing the difference for each $B-A_j$ as a rotation around the X axis with controlled gates $R_x$.
The gate provides a rotation only if the values of the qubits with the same significance are not equal to each other. That is, if you compare states $101$ and $011$, the qubit rotation will be at the expense of the high and middle bits, since the values of the low bits coincide and are equal to $1$. The direction of rotation is initialized in any convenient way, only the sign of the difference determines the change of direction: if at a negative result the rotation is chosen clockwise, then at a positive one must be chosen counter-clockwise, and vice versa. This is necessary for the correct calculation of the difference, so $101-011 = 010$, which in this example is equivalent to $+ \frac{\pi}{2} - \frac{\pi}{4} = \frac{\pi}{4}$. If all turns are made in the same direction, the angle will be equal to $3\frac{\pi}{4}$, which is equivalent to $110$, and this is incorrect.
Here we offer only one of the many options for the dependence of the rotation angle on the number of bits. For example, you can turn the highest bit by $2\frac{\pi}{3}$ and the next by $\frac{\pi}{3}$, and no more rotations. The values can be calibrated depending on the expected effect on the final result, especially if something is known about the data (for example, that they are no more or less than a certain value, or it is necessary to weaken all values to a certain threshold). In this case, the bitwise setting can take any convenient form. The only restriction is that the sum of all rotations does not exceed $\pi$, otherwise the reverse changing in amplitudes may begin, as in the case if you make unnecessary oracle calls in Grover's algorithm.
Thus, if the state $A_j$ is entangled with the state of a special $D$ qubit was $|0\rangle$ with each turn the probability of the value $|1\rangle$ will increase in proportion to the difference $B-A_j$. It does not matter whether the total (by all digits) turn occurs clockwise or counterclockwise (more is subtracted from less or vice versa), because eventually in both cases the probability will "move away" from $|0\rangle$ and get closer to $|1\rangle$. Consequently, the need to consider the sign is discarded.
The following values will be taken as an example: $|B\rangle=|101\rangle$,  $|A_0\rangle=|010\rangle$,$|A_1\rangle=|110\rangle$; $B-A_0=101 - 010= 011$.
The result $011$ should be written in qubit $D$ as a rotation around the $X$-axis, the abstraction of the amplitude amplification $|A_1\rangle|1\rangle$ is shown in Figure~\ref{fig:4RotExmp}.
In calculating $B-A_1=101-110=-001$ the difference is negative, which does not affect the absolute value of the probability amplitude of state $|A_0\rangle|0\rangle$ corresponding to a rotation angle of $\frac{\pi}{4}$ (Figure~\ref{fig:4RotExmp}).
As you can see, there is no need to consider the sign of the difference, because in this case it only affects the direction of motion: clockwise or counterclockwise ($+\frac{\pi}{2}$ or $-\frac{\pi}{2}$). $101-010$ will be equal to $011-110$: the probability of getting $|0\rangle$ when measuring the counter $D$ will decrease towards $|1\rangle$ in both cases equally.
So, given $|B\rangle=|101\rangle$, $|A_0\rangle=|010\rangle$, $|A_1\rangle=|110\rangle$ the operation of the circuit shown in Figure~\ref{fig:4RotExmp} is described by equation:
\begin{equation}             \label{eq8}
\begin{gathered}
|B\rangle\left|A_0\right\rangle\left|A_1\right\rangle|000\rangle|0\rangle \xrightarrow{H} \\
\xrightarrow[\rightarrow]{H} \frac{1}{\sqrt{2}}|B\rangle\left|A_0\right\rangle\left|A_1\right\rangle|000\rangle(|0\rangle+|1\rangle) \xrightarrow{I} \\
\xrightarrow{I} \frac{1}{\sqrt{2}}|B\rangle\left|A_0\right\rangle\left|A_1\right\rangle\left(\left|A_0\right\rangle|0\rangle+\left|A_1\right\rangle|1\rangle\right)= \\
=\frac{1}{\sqrt{2}}|101\rangle|010\rangle|110\rangle(|010\rangle|0\rangle+|110\rangle|1\rangle) \xrightarrow{I I} \\
\quad \xrightarrow{I I} \frac{1}{\sqrt{2}}|101\rangle|010\rangle|110\rangle \otimes \\
\otimes\left[| 0 1 0 \rangle \left(\operatorname { c o s } \frac { \pi } { 4 } \left(\cos \frac{\pi}{8}\left(\cos \frac{\pi}{16}|0\rangle-i \cdot \sin \frac{\pi}{16}|1\rangle\right)+\right.\right.\right. \\
\left.+i \cdot \sin \frac{\pi}{8}\left(-i \cdot \sin \frac{\pi}{16}|0\rangle+\cos \frac{\pi}{16}|1\rangle\right)\right)- \\
-i \cdot \sin \frac{\pi}{4}\left(i \cdot \sin \frac{\pi}{8}\left(\cos \frac{\pi}{16}|0\rangle-i \cdot \sin \frac{\pi}{16}|1\rangle\right)+\right. \\
\left.\left.+\cos \frac{\pi}{8}\left(-i \cdot \sin \frac{\pi}{16}|0\rangle+\cos \frac{\pi}{16}|1\rangle\right)\right)\right)+ \\
+|110\rangle\left(i \cdot \sin \frac{\pi}{8}\left(\cos \frac{\pi}{16}|0\rangle-i \cdot \sin \frac{\pi}{16}|1\rangle\right)+\right. \\
\left.\left.+\cos \frac{\pi}{8}\left(-i \cdot \sin \frac{\pi}{16}|0\rangle+\cos \frac{\pi}{16}|1\rangle\right)\right)\right]
\end{gathered}
\end{equation}
\section{Conclusion}
Taking into account all the arithmetic calculations after measuring the counter $D$ the probability of getting $|0\rangle$ is $~37\%$ and probability of getting $|1\rangle$ is $~63\%$. So, with a probability of $~63\%$ depending on the proportions of differences $B-A_0$ and $B-A_1$, the nearest to $B$ value will be determined by the element $A_1$, which means that the goal by using the proposed scheme is achieved.
This algorithm can be considered as an anti-search, which may not sufficiently enhance the desired result, but significantly limit the probability of measuring "bad" values. Or use it as a filter that allows you to influence the output value in one way or another. At the same time, by calibrating the angle of rotation, the degree of this influence can be varied. It should be noted that the gains may be insufficient and more likely to be advisory in nature (especially as in the case of the example where the difference between the two elements is not so great), and the algorithm should be used in tasks where high accuracy is not needed, but but overall a good result and high processing speed are desirable.
The implementation options for the proposed rotation matrix, as well as iterativity, i.e. the possibility of further increasing the amplitude with repeated oracle calls, are an independent tasks and a reason for discussion at this stage of the study.

\section*{Acknowledgements}
I want to express my gratitude to the free application 
$``$Quantum Computing$''$ by $hex@dec$:
\\
\url{https://play.google.com/store/apps/details?id=hu .hexadecimal.quantum&hl=en_AU&gl=US}
\\This application allowed me to perform verification calculations to make sure that my algorithm works.
I also want to express my gratitude to Yuri Ozhigov for his basic knowledge in this field.

\end{document}